\documentstyle[aps,multicol,prl]{revtex} 
\begin{document}
\title{ 
Interfaces of modulated phases
}
\author{
Roland R. Netz$^{*,\$}$,
 David Andelman$^{\#}$, 
and M. Schick$^*$
}
\address{$^*$Department of Physics, Box 351560, University 
of Washington, Seattle, WA 98195-1560}
\address{$^{\$}$Max-Planck-Institut f\"ur Kolloid- und Grenzfl\"achenforschung,
Kantstr. 55, 14513 Teltow, Germany}
\address{$^{\#}$School of Physics and Astronomy,
Tel-Aviv University, Ramat Aviv 69978, Tel Aviv, Israel}

\date{\today}
\maketitle
 
\begin{abstract} 
Numerically minimizing a continuous free-energy functional which yields 
several modulated phases, we obtain the order-parameter profiles and 
interfacial free energies of symmetric and non-symmetric tilt boundaries 
within the lamellar phase, and of interfaces between coexisting lamellar, 
hexagonal, and disordered phases.  Our findings agree well with chevron, 
omega, and T-junction tilt-boundary morphologies observed in diblock 
copolymers and magnetic garnet films.

 \end{abstract}
 

\begin{multicols}{2} 
\narrowtext          

Modulated phases are found in a surprisingly diverse set of physical and
chemical systems, including superconductors, thin-film magnetic garnets
and ferrofluids, Langmuir monolayers, and diblock
copolymers\cite{Science}.  Such phases are characterized by periodic
spatial variations of the pertinent order parameter in the form of
lamellae, cylinders, or cubic arrangements of spheres or interwoven
sheets.  This self-organization results from competing interactions: a
short-ranged molecular one favoring a homogeneous state, and a
long-ranged contribution, which can have magnetic, electric, or elastic
origin, favoring domains.  Because of the modulation, interfaces between
different phases or grain boundaries within a single phase are most
interesting, yet they have received much less attention than those
occurring in solids.  Recently, experimental
studies of grain boundaries within lamellar phases of diblock copolymer
have been carried out\cite{Thomas,Hashimoto} which illustrate the rich
interfacial behavior exhibited by such systems.  In particular, three
morphologies of tilt boundaries[2b], denoted chevron, omega, and
T-junction, were observed.  Such interfaces are more difficult to
describe theoretically than those between uniform phases, which have
been the subject of classic work\cite{Helfand}. 
In this article we study not only tilt grain
boundaries within lamellar phases, but also interfaces between
coexisting modulated phases of different symmetry.

The dimensionless free-energy functional we use,
\begin{eqnarray}
{\cal F}[\phi]&=&
\int \left\{ -\frac{\chi}{2} \phi^2 
+\frac{1-\phi}{2} \ln \frac{1-\phi}{2}
+\frac{1+\phi}{2} \ln \frac{1+\phi}{2} \right. \nonumber \\
&& \left. -\frac{1}{2}(\nabla \phi)^2 +
\frac{1}{2}(\nabla^2 \phi)^2 - \mu \phi \right\}dV ,
\end{eqnarray}
includes an enthalpic term (proportional to the interaction parameter
$\chi$), favoring an ordered state in which $|\phi|$ is non-zero, an
entropy of mixing preferring a disordered state, $\phi=0$, and
confining $|\phi|$ to be less than unity, and derivatives of the order
parameter.  The ordered state occurs with a modulation of dominant wave
vector $q^*=1/\sqrt{2}$ because of the competition between the negative
gradient square term (favoring domains at large length scales) 
and the positive Laplacian square (preferring a homogeneous state at 
small length scales).
Such a free energy functional has been used to describe the bulk phases
of magnetic layers\cite{Garel}, Langmuir films\cite{Andelman},
amphiphilic systems,\cite{Gompper2}, and diblock
copolymers\cite{Leibler1980}, and the effects of surfaces on
isotropic\cite{Fredrickson} and hexagonal phases\cite{Turner} of the
latter.

In our numerical studies, we determine the minimum of (1) {\em directly}
using a conjugate-gradient method, which is more convenient than solving
the corresponding Euler-Lagrange equation.  We discretize $\phi$ on a
square lattice, approximate the derivatives by nearest-neighbor
differences, and employ a mesh size sufficiently small that
discretization effects are negligible within the numerical precision. A
typical grid contains 40,000 points.  Fig.1 shows the resulting
two-dimensional bulk phase diagram as a function of $\chi$ and a) the
average order parameter $\Phi\equiv \langle \phi\rangle_V$ and b) the
chemical potential $\mu$. It is in good agreement with previous
calculations based on single-mode
approximations\cite{Garel,Andelman}. In addition to the disordered (D)
phase, there is a lamellar (L) and two hexagonal (H) phases, which all
join at the critical point at $\chi_C=3/4$, $\mu=\Phi=0$.  (In three
dimensions, one expects additional cubic phases, which are not studied
here.) For larger values of $\chi$, the H and L phases each terminate at
a triple point, whereas experimentally one sees that these modulated
phases exist even for very large values of $\chi$.  This unphysical part
of the phase diagram is due to a breakdown of the gradient expansion in
(1).

We first present results for grain boundaries in lamellar phases,
and begin with the asymmetric tilt grain boundary 
(GB) between two perpendicular L phases.
This is the {\em T-junction} of Ref. [2b], for which the layer
continuity between the two adjoining phases is disrupted. 
Figs.2a-c show order-parameter profiles $\phi$
for different $\chi$ and $\mu=0$ (symmetric stripes), 
while Figs.2d-f show profiles for varying
chemical potential $\mu$ and
$\chi=1$ (asymmetric stripes). 
Fig.2b clearly shows the enlarged endcaps noted in
experiment\cite{Science,Thomas}, and the series 2a-c predicts that they
become less pronounced with increasing $\chi$.  The GB interfacial  energy
$\gamma_{GB}$ scales as $\gamma_{GB} \sim (\chi-\chi_C)^{\mu_*}$ with
$\mu_*=3/2$, Fig.2g, in accord with mean-field
predictions\cite{Rowlinson}.  To demonstrate that the interfacial energy
$\gamma_{GB}$ is indeed quite small, we show with a broken line the
difference in bulk free energies between the disordered and lamellar
phases multiplied by one wavelength of the latter. This is roughly the
cost per unit area of disordering such a width of lamellar
phase. Clearly the actual grain boundary bridges the two grains in a
manner much less expensive than the insertion of a region of disorder.
In Fig.2h we plot the interfacial energies of two distinct GB
structures, which are degenerate at $\mu=0$, being related by
order-parameter reflection symmetry $\phi \rightarrow -\phi$. For $\mu
\neq 0$ this symmetry is broken and the two structures correspond to
distinct local free energy minima, one metastable, the other globally 
stable. The free-energy barriers between such 
interfacial structures are responsible for slow interface
motion and thus long healing times in
multi-grain lamellar samples\cite{Cahn}.  

In Figs.3a-c we show
symmetric tilt-boundary (TB) configurations for a fixed angle
$\theta=90^\circ$ between the layer normals as a function of $\chi$. 
Here the layer continuity is maintained across the boundary, and
the {\em chevron} morphology is quite evident. In Figs.3d-f, the tilt angle is
progressively increased at fixed $\chi$, and one clearly sees the change
from the chevron to the {\em omega} structure.
The omega shape of the layers at the boundary results from
frustration due to an imposed  local lamellar wavelength
much larger than the equilibrium value.
Figs.3d-f resemble micrographs of
undulating lamellar patterns in garnet films\cite{Seul2}, and the
similarity of Fig.3f to the micrographs[2b] of diblock
copolymer TB's is striking.  The scaling of the TB energy $\gamma_{TB}$
is again described by an exponent $\mu_*=3/2$ (Fig.3g).  Close to
criticality, one finds pronounced reconstruction in terms of a
square-like modulation, Fig.3a.  For small $\theta$,
$\gamma_{TB}(\theta) \sim \theta^3$, Fig.3h, in accord with the bending
behavior of elastic sheets; the $\theta \rightarrow 180^\circ$ limit is
expected to be linear, $ \gamma_{TB}(\theta) \sim 180^\circ-\theta$, in
accord with a description in terms of decoupled dislocations of finite
creation energy.

We now turn to interfaces between thermodynamically distinct phases.
Interfacial structures between coexisting lamellar and disordered phases
for three different values of the angle $\vartheta$ between the
lamellae and the interface are shown in Figs.4a-c for $\chi=1$
(where the phases and, therefore, also the interfacial structure
are {\em metastable}), and
in 4d-f for $\chi=1.5$ (where they are stable). 
The metastable L-D boundary
is shown as a dashed line in Fig.1b.
In  Figs.4a and d,
for $\vartheta=0^{\circ}$, there is a
relaxation of the outermost layers as the interface is approached,
leading to a small increase in the wavelength, as in solids.

Before presenting the interfacial free energies, we describe our method
of obtaining them, because the calculation of such free energies 
between coexisting phases
which can be modulated 
is non-trivial\cite{Widom2}.
We calculate the total free energy, $F^I$,
of phase $I$ in a box employing periodic boundary conditions parallel to 
the left and right faces, and 
reflecting (Dirichlet) boundary
conditions on those faces themselves. We adjust the length of the box so
that the free energy is minimized. This occurs when the length of the
box is some integer number of wavelengths of the periodic structure of
phase $I$. The volume of the box is $V^\prime$.  By this means, there is
{\em no} surface contribution to the total free energy, so that the bulk
free energy is obtained directly; $f_b^I=F^I(V^{\prime})/V^{\prime}.$ In
a similar way we obtain
$f_b^{II}=F^{II}(V^{\prime\prime})/V^{\prime\prime}.$ For the system in
$I,\ II$ coexistence, we calculate the total free energy in a box large enough
that the order parameters attain their bulk values on the left and right
faces at which reflecting boundary conditions are employed, and we vary
the length of the box to minimize the total free energy. The volume of
the box is $V$. Again there are no surface contributions so that we
obtain $F^{I,II}(V,A)=V(f_b^I+f_b^{II})/2+A\gamma_{I,II}.$ As the bulk
free energies are known, the desired interfacial energy follows
directly.

The interfacial energy $\gamma_{LD}$ is shown in Fig.4g. 
As $\chi$ is decreased, the coexistence between L
and D is preempted by the hexagonal phase (see Fig.1b). This is
manifest in the behavior of $\gamma_{LD}$ for
$\vartheta=90^\circ$ (broken line in Fig.4g) which becomes negative.
This value of $\gamma_{LD}$ is obtained by assuming an interfacial
reconstruction locally resembling the energetically preferred hexagonal
phase (see Fig.4c).  The value of $\gamma_{LD}$ for $\vartheta=0^\circ$ 
(solid line in Fig.4g) remains
positive down to  $\chi\simeq 0.82$  and vanishes
there with the classical tricritical exponent $\mu_*=2$\cite{Widom1}.
Strong anisotropy of $\gamma_{LD}$ as function of $\vartheta$ is
observed (Fig.4h), from which the shape of lamellar droplets in an
isotropic phase can be qualitatively inferred: For $\chi> 1.34$ 
the drops are elongated parallel to the lamellae;
for $\chi<1.34$ the L and D phases are metastable at
coexistence and the drops are elongated perpendicular to the lamellae.

Hexagonal-disordered interfacial structures are de\-pic\-ted in Figs.5a-c; 
the corresponding interfacial energy $\gamma_{HD}$, shown in Fig.5g, scales
again with $\mu_*=2$.  Structures for the
hexagonal-lamellar interface are plotted in Figs.5d-f, where we chose
the mutual orientation in accord with the epitaxial relationship found
for amphiphilic systems\cite{Rancon}.  
The corrugation of the lamellae near the
interface, particularly evident in Fig.5d, resembles that seen in
experiments on diblock copolymer blends\cite{Spontak}.
The interfacial free energy of
this interface, $\gamma_{HL}$, is plotted in Fig.5h and scales with the 
classical critical  exponent $\mu_*=3/2$.
This reflects the fact that, in contrast to the L-D and H-D interfaces,
the H and L  phases are locked into
a fixed relative position with respect to translations perpendicular
to the H-L interface.

At the triple point between disordered, lamellar, and hexagonal phases,
$\chi \simeq 1.34$, we find
$\gamma_{LD} >\gamma_{HL}+\gamma_{HD}$. It follows that the L-D
interface that we have calculated is not the
thermodynamically stable one. Within our model, the L-D
interface is therefore wetted by the hexagonal phase at the triple
point. As the occurrence of such points between these three phases is
not uncommon experimentally\cite{Strey}, it would be interesting to
determine whether this wetting does indeed occur.

Density fluctuations will change the interfacial critical exponent from
$\mu_*=3/2$ to $\mu_*=1.26$, but leave the classical tricritical
exponent $\mu_*=2$ intact\cite{Rowlinson}, 
while fluctuations of the direction of the
modulation normals will eliminate the critical
point\cite{Brazovskii}.  By introducing uniaxiality, these latter
fluctuations can be suppressed.

In summary, we have employed a simple Ginzburg-Landau 
free-energy functional to
calculate profiles and free energies of several interfaces of modulated
phases.  Qualitative agreement with experiment is very good.  The
observed chevron and omega morphologies at lamellar
grain boundaries emerge naturally, as do the
expanded endcaps characteristic of the T-junction.  We also calculated
profiles and free energies of interfaces between disordered and
modulated phases, and between modulated phases of different symmetry. In
all but the simplest cases, there is significant reconstruction which
leads to low interfacial energies. An extreme example of such
reconstruction is the  lamellar-disordered interface,
which we find to be wetted by the hexagonal phase at 
an L-H-D triple point.

This work was supported in part by grants from the United States-Israel
Binational Science Foundation under grant
94-00291, and the National Science Foundation,
grant DMR9531161. We thank Ned Thomas for bringing the work of
Gido et al. to our attention.

\begin{figure}
\caption{
Two-dimensional bulk phase diagram, showing disordered (D), lamellar
(L), and hexagonal (H) phases, as a function of the interaction
strength $\chi$ and a) the average order parameter 
 $\Phi$ and b) the chemical
potential $\mu$. Broken lines in a) denote triple lines, broken lines in
b) denote the (metastable) L-D transitions which exhibit tricritical points
(denoted by filled circles).}

\caption{Grain boundary (GB) between two perpendicular lamellar phases.
a-c) Contour plots of the order parameter profiles
 for $\mu=0$ and $\chi=0.78$, $1$, $1.5$; throughout the article
 the order parameter range $[-1,1]$ is represented by 20 grayscales; d-f)
Profiles for $\chi=1$ and $\mu=-0.02$, $ 0.02$, $0.06$.  g) GB
interfacial energy $\gamma_{GB}$ for $\mu=0$ as a function of $\chi$,
showing asymptotic scaling with an exponent $\mu_*=3/2$ (inset). The
dashed line gives the surface free energy density of one lamellar layer
for comparison.  h) $\gamma_{GB}$ for $\chi=1$ as a function of $\mu$
for two GB configurations which are degenerate at $\mu=0$, but distinct
otherwise.}

\caption{Tilt boundary (TB) between two lamellar  phases.
a-c) Profiles for $\mu=0$, a fixed angle $\theta=90^\circ$ between the
layer normals  and $\chi=0.78$, $1$, $1.5$; d-f) Profiles for
$\mu=0$, $\chi=1$, and $\theta=28.08^\circ$, $53.14^\circ$, and
$126.86^\circ$.  g) TB interfacial energy $\gamma_{TB}$ for $\mu=0$ as a
function of $\chi$, showing asymptotic scaling with an exponent
$\mu_*=3/2$ (inset).  h) $\gamma_{TB}$ for $\chi=1$ as a function of
$\theta$, showing a $\gamma_{TB} \sim \theta ^3$ behavior for small
$\theta$ (broken line) and  linear behavior for 
$\theta \rightarrow 180^\circ$ .}

\caption{Lamellar-disordered  interface.
a-c) Profiles for $\chi=1$ and different angles $\vartheta=0^\circ$,
$45^\circ$, and $90^\circ$ between the
lamellae and the interface; d-f) profiles for $\chi=1.5$.  g) Interfacial
energy $\gamma_{LD}$ for $\vartheta=90^\circ$ (broken line) and
$\vartheta=0^\circ$ (solid line), the latter scaling with $\mu_*=2$ on
approach to the metastable tricritical point (inset).  h) $\gamma_{LD}$
as a function of the angle $\vartheta$ for $\chi=1$ (broken line) and
for $\chi=1.5$ (solid line).}

\caption{
a-c) Hexagonal-disordered interfacial profiles for $\chi=0.78$, $0.9$,
and $1.2$; d-f) Hexagonal-lamellar profiles for $\chi=0.78$, $0.9$, and
$1.2$.  g) Interfacial energy $\gamma_{HD}$ along HD coexistence as a
function of $\chi$, scaling with $\mu_*=2$ (inset); h) $\gamma_{HL}$ as
a function of $\chi$, scaling with $\mu_*=3/2$ (inset).}

\end{figure}
\end{multicols} 

\begin{references} 
\bibitem{Science}
M. Seul and D. Andelman, Science {\bf 267}, 476 (1995).

\bibitem{Thomas}
a) S.P. Gido, J. Gunther, E.L. Thomas, and D. Hoffman, Macromolecules {\bf
26}, 4506 (1993); b) S. P. Gido and E.L. Thomas, Macromolecules {\bf 27},
6137 (1994).

\bibitem{Hashimoto}
Y. Nishikawa {\em et al.},
Acta Polymer. {\bf 44}, 192 (1993);
T. Hashimoto, S. Koi\-zu\-mi, and H. Hasegawa, Macromolecules
{\bf 27}, 1562 (1994).

\bibitem{Helfand}
E. Helfand and Y. Tagami, J. Chem. Phys. {\bf 56}, 3592 (1972);
L. Leibler, Macromolecules {\bf 15}, 1283 (1982);
A.N. Semenov, Sov. Phys. JETP {\bf 61}, 733 (1985).

\bibitem{Garel}
T. Garel and S. Doniach, Phys. Rev. B {\bf 26},325 (1982).

\bibitem{Andelman}
D. Andelman, F. Brochard, and J.-F. Joanny, J. Chem. Phys.
{\bf 86}, 3673 (1987).

\bibitem{Gompper2} 
G. Gompper and M. Schick, Phys. Rev. Lett. {\bf 65}, 1116 (1990).

\bibitem{Leibler1980} L. Leibler, Macromolecules {\bf 13}, 1602 (1980);
G.H. Fredrickson and E. Helfand, J. Chem. Phys. {\bf 87}, 697 (1987).

\bibitem{Fredrickson}
G.H. Fredrickson, Macromolecules {\bf 20}, 2535 (1987).

\bibitem{Turner}
M.S. Turner, M. Rubinstein, and C.M. Marques,
Macromolecules {\bf 27}, 4986 (1994).

\bibitem{Rowlinson} 
J.S. Rowlinson and B. Widom, {\em Molecular Theory of Capillarity} 
(Oxford University Press, New York, 1982). 

\bibitem{Cahn}
Similar local free-energy minima for a moving interface had been postulated
in the context of crystal growth,
J.W. Cahn, Acta Met. {\bf 8}, 554 (1960).

\bibitem{Seul2}
M. Seul and R. Wolfe, Phys. Rev. Lett. {\bf 68}, 2460 (1992);
Phys. Rev. A {\bf 46}, 7519 (1992).

\bibitem{Widom2}
This problem has been formulated within the context of
coexisting lamellar phases,
A. Chatterjee and B. Widom, Mol. Phys. {\bf 80}, 741 (1993);
B. Widom, KNAW Symposium (Amsterdam, 1994).

\bibitem{Widom1}
B. Widom, Phys. Rev. Lett. {\bf 34}, 999 (1975); see also
 Ch.9.5 of \cite{Rowlinson}.

\bibitem{Rancon}
Y. Rancon and J. Charvolin, J. Phys. Chem. {\bf 92}, 6339 (1988);
K.M. McGrath, P. K\'ekicheff, and M. K\'leman, J. Phys. II France
{\bf 3}, 903 (1993).

\bibitem{Spontak}
R.J. Spontak {\it et al.}, Macromolecules {\bf 29}, 4494 (1996).

\bibitem{Strey} For an example, see R. Strey, Ber. Bunsenges. Phys. Chem. 
{\bf 97}, 742 (1993).

\bibitem{Brazovskii}
S.A. Brazovskii, Sov. Phys. JETP {\bf 41}, 85 (1975). 
\end{references}
\end{document}